# Modelling the Strategic Alignment of Software Requirements using Goal Graphs

Richard Ellis-Braithwaite[1]   Russell Lock[1]   Ray Dawson[1]   Badr Haque[2]
[1]Loughborough University   [2]Rolls-Royce Plc.
Leicestershire, United Kingdom   Derby, United Kingdom
{r.d.j.ellis-braithwaite@lboro.ac.uk, r.lock@lboro.ac.uk, r.j.dawson@lboro.ac.uk, badr.haque@rolls-royce.com}

*Abstract*—**This paper builds on existing Goal Oriented Requirements Engineering (GORE) research by presenting a methodology with a supporting tool for analysing and demonstrating the alignment between software requirements and business objectives. Current GORE methodologies can be used to relate business goals to software goals through goal abstraction in goal graphs. However, we argue that unless the extent of goal-goal contribution is quantified with verifiable metrics and confidence levels, goal graphs are not sufficient for demonstrating the strategic alignment of software requirements. We introduce our methodology using an example software project from Rolls-Royce. We conclude that our methodology can improve requirements by making the relationships to business problems explicit, thereby disambiguating a requirement's underlying purpose and value.**

## I. INTRODUCTION

The stakeholders of a software project should share an understanding of the potential business benefit that a software requirement offers. If such an understanding can be achieved, the likelihood that a solution will satisfy a real business problem will be improved. Although such statements may sound obvious, it has been reported that 45% of software requirements are never deemed to be useful after implementation [1]. These unnecessary requirements cause costs and delays that perhaps could have been avoided by benefit analysis. The existence of a requirement should be questioned if it does not demonstrate potential to offer value to the business. Conversely, valuable requirements are at risk of being de-prioritised if they fail to demonstrate their potential benefit. In an organisational setting, business benefit can be gained from an alignment to business strategy.

Technically worded requirements or solution oriented requirements (i.e., specified for the machine rather than for the world [2]) hide the business problem to be solved and leave stakeholders with little understanding of the potential value. It is therefore important that the strategic alignment of such a requirement is explored in order to avoid wastage.

This paper explores the suitability of goal graphs for demonstrating a software requirement's strategic alignment. Current Goal Oriented Requirements Engineering (GORE) standards, such as GRL [3], do not quantify the contribution one goal makes to another using metrics from the application domain, opting instead to use scales such as high, medium and low, or numerical scales such as 0-9. As a result, any strategic alignment proposed by the use of goal graphs is not specific, measurable or testable. Proposed extensions by Van Lamsweerde [4] do not consider that a chain of linked goals may contain a variety of metrics that need to be translated in order to demonstrate strategic alignment. Additionally, the current methods do not consider how the contribution score is calculated and how that affects the credibility and accuracy of the proposed benefits. This paper attempts to demonstrate how the above problems can be addressed, thereby allowing goal graphs to be used to analyse the strategic alignment of software requirements. Our methodology complements frameworks that require business value analysis, such as value-based software engineering [5], by making assumptions about business value explicit.

We have developed and implemented our methodology in partnership with an industrial partner (Rolls-Royce) to ensure its utility in real world settings. We use examples in the context of a software project to be implemented in the Transmissions Structures & Drives (TS&D) Supply Chain Unit (SCU). The software will automate geometry design and analysis for aero engine components, as well as for their manufacturing tools such as casting molds. Simply put, engineers will input the desired design parameters and the software will output the component's geometry.

In Section II, we introduce the problem that this paper addresses, while in Section III, we present and evaluate the extent to which existing solutions address it. Section IV presents our methodology and tool as an extension of an existing GORE methodology in order to address the gaps outlined in Section III. We conclude in Section V with remarks on the paper's contributions and future work.

## II. THE PROBLEM

Ross and Schoman stated that software requirements "must say why a system is needed, based on current or foreseen conditions" as well as "what system features will serve and satisfy this context" [6]. Popular Requirements Engineering meta models [7], [8] and templates [9], [10] tend not to focus on "why", typically addressing it by stipulating that rationale be attached to a requirement. However, rationale is not always an adequate description of why the requirement is valuable to the business. If only one "why" question is asked about the requirement then the rationale can still be distant from the true problem to be solved (i.e., the essence of the requirement), and it may be defined without consideration of its wider implications.

As an example of the problem that this paper examines, we introduce the following high-level requirement taken from our example software project: "While operating in an analysis solution domain and when demanded, the system shall run analysis models". The business value of this software requirement and the underlying problem to be solved is not immediately obvious, so to better understand the need for this requirement, we examine the attached rationale: "So that structural integrity analysis models can be solved as part of an automated process". The requirement's benefit to the business and its alignment with strategy are still not clear after one level of abstraction above the requirement. Additionally, the extent of the problem to be solved is not explained, i.e., the problems associated with solving structural integrity analysis models manually and the wider implications of doing so. Perhaps the manual process is costing the business in terms of human resource time or inaccuracy of the analysis due to error. If so, what is the business impact and how far can it be reduced? Additional broader implications may exist that are not immediately obvious - it might be the case that design innovation is constrained by the slower manual process. Clearly there is more to the rationale than is written, and arguably more than it would be sensible to express within a requirement, partly due to the duplication this would incur; several requirements may achieve the same business benefit but at varying levels of contribution and with potentially complex dynamics.

In summary, this paper argues that the strategic alignment of a requirement should be examined so that:
1. The root of the requirement can be understood so that the software can solve the right problem.
2. The extent of the problem can be understood to prove the requirement's value and validity.
3. The value of the requirement can be understood to better inform prioritisation and project funding.

### III. BACKGROUND

The following areas of research are related to the strategic alignment of software requirements: (A) Goal Oriented Requirements Engineering, (B) Strategic Alignment and (C) Software Metrics.

#### A. Goal Oriented Requirements Engineering

Goal Oriented Requirements Engineering (GORE) seeks to provide answers to the, so far, largely unanswered question of "why" software functionality should exist through the use of goal graphs. Van Lamsweerde defines the term goal in the context of GORE as an optative statement (i.e., desired future state) about an objective that the system hopes to achieve [11]. "Goal" in the context of GORE is therefore more concerned with the goals of the system than the goals of the business. Furthermore, this definition of goal does not differentiate it from an objective. In order to relate the goals of the system to the goals of the business, we need an integrated definition of the terms used in business strategy. The Object Management Group (OMG) defines these terms in the Business Motivation Model (BMM) [12]. The BMM defines a goal as an indication of "what must be satisfied on a continuing basis to effectively attain the vision of the business". An objective is then defined as a "statement of an attainable, time-targeted, and measurable target that the enterprise seeks to meet in order to achieve its goals".

Objectives therefore contrast with goals in that "goals are allowed to be unrealistic and unachievable" [13]. Attempting to prove strategic alignment to non-specific goals such as "maximise profit" would be difficult since it would not be possible to prove the extent of its satisfaction. Therefore, requirements should be abstracted to objectives rather than goals for strategic alignment. Fortunately, business strategies are usually decomposed into objectives that follow SMART [14], which allows contribution between objectives to be specified, e.g., "objective x will satisfy half of objective y".

Since the only significant difference between an objective and a goal is in its hardness and specificity (i.e., whether its satisfaction can be determined), GORE methodologies can still be applied. The most well-known GORE methodologies include KAOS [15], i* [16] and GRL [17]. Such methodologies produce goal graphs whereby goals at a high level represent the end state that should be achieved and lower level goals represent the means to that end. The relationships between goals are typically expressed as means-ends links with AND/OR refinement. Additional elements such as agents, obstacles and dependencies are typically included. A goal graph is traversed upwards in order to understand why a goal should be satisfied and downwards to understand how that goal could be satisfied.

Three methods for applying weights to goal-goal contribution links in goals graphs were proposed by Van Lamsweerde [4] with the intention of extending KAOS, but the concepts could be applied to any GORE method:
1. Subjective qualitative scores e.g., --, -, +, ++.
2. Subjective quantified scores e.g., 0 to 100.
3. Objective gauge variables (i.e., a quantity prescribed by a leaf soft goal to be increased, reduced, etc.).

After evaluating the above options, Van Lamsweerde concluded that the specification of link weight scores with objective gauge variables is the most appropriate approach due to its verifiability. Van Lamsweerde goes on to demonstrate how alternative goal (or requirement) options can be evaluated by estimating the contribution a goal makes to soft goals. Soft goals are typically qualitatively stated desires used for comparing alternatives, but in the context of [4], fit criterions quantify them. A number of translations between soft goals often need to be made in order to link requirements to business goals, which inevitably involves translating metrics (e.g., reducing component design *time* contributes to reducing component *costs*). However, the method presented in [4] does not provide guidance on propagating contribution scores to high level (i.e., abstracted) soft goals. This is important since a requirement may contribute positively to system level goals, but only slightly to higher business goals. The only propagation approach prescribed is cumulative addition (a goal's contribution = the sum of the contributions made by the goal's children), which cannot be applied because two scales cannot be summed.

Goal Requirements Language (GRL) was recently made an international standard through ITU specification Z.151 as part of the User Requirements Notation (URN) [3]. GRL integrates the core concepts of i* and NFR [17]. Link contributions in GRL are specified with subjective quantified contribution scores, much the same as outlined in Van

Lamsweerde's paper [4]. For example, the time reduction goal might contribute to the cost saving goal with a contribution weight of 67 out of 100. This contribution score is untestable and meaningless; moreover it is not refutable, which, according to Jackson [2], means that the relationship is not described precisely enough because no one can dispute it. The only way such scales could be meaningful is if the goals were specified with fit criterions (e.g., a cost to be saved) and if the scales implied percentage satisfaction (which they do not). In which case, a 50/100 contribution might imply that 50% of a £20,000 annual cost saving will be achieved. However, this is only applicable for goals whose satisfaction upper bound is 100%, since the scale's upper bound is 100; which is not the case for goals involving increases, which may specify more than 100%. Recently, the jUCMNav tool allowed for the relation of Key Performance Indicators (KPIs) to goals in GRL [18]. KPIs specify business targets that measure the performance of a business activity. However, since KPIs do not affect the way in which contribution is measured (i.e., the contribution that a chain of goals makes to a KPI), subjectivity and ambiguity still exists.

One of the most popular tools to compare product qualities with customer requirements is the House of Quality (HoQ) diagram [19]. The fundamental failing of the HoQ is that the score values used to measure the strength of the contribution are subjective, much like those used in GRL. Additionally, since the HoQ is constructed using a 2D grid, only two dimensions can be compared in the same grid, i.e., requirements can be related to software goals, but if those software goals are to be related to customer or business goals, then additional grids will be required for each extra dimension. If these dimensions are not explored (e.g., if the software project goals are not abstracted to business goals), then the goals that the alternative solutions will be evaluated against may be incorrect (e.g., solution specific or aiming to solve the wrong problem). GORE methodologies which evaluate alternative solutions against their effect on goals, e.g., [20], also depend on the alignment of those goals to higher level goals for the resulting decision to be correct.

*B. Strategic Alignment*

One of the most suitable methodologies for relating software requirements to business strategy is B-SCP [21], due to its tight integration with the OMG's Business Motivation Model (BMM). B-SCP decomposes business strategy towards organisational IT requirements through the various levels of the BMM (e.g., the vision, mission, objective, etc.). However, B-SCP cannot accurately show that a requirement satisfies a strategy since no contribution strengths are assigned to links. Since strategic alignment depends on the extent to which the strategy is satisfied (e.g., for the goal "reduce costs", the extent is the amount of cost to be reduced), the extent to which a goal contributes towards another needs to be considered. Indeed, a large proportion of software requirements will only partially satisfy the strategic objectives. Moreover, B-SCP's methodology refines business strategy towards IT requirements, which means that completeness of the model is dependent on the completeness of the business strategy, i.e., there is no opportunity to refine software functionality upwards to propose new business strategy. Additionally, B-SCP does not consider goal conflicts, dependencies, actors or obstacles, as in the GRL and KAOS methodologies.

The Balanced Scorecard and Strategy Maps [22] approach offers guidance on formulating and relating business goals to each other under four perspectives: financial, customer, internal processes, learning and growth. The approach does not concern software requirements; but such an approach could be performed before software requirement to business strategy alignment analysis takes place, in order to ensure business strategy completeness.

*C. Metrics*

Fit criterions as specified by Volere [9] and Planguage [23] can be attached to requirements in order to make them measurably satisfiable. However, assumptions made about the benefits that may be reaped after satisfaction of a fit criterion are not addressed in either methodology. Additionally, Volere and Planguage propose textual representation of requirements; and as such, relationships between requirements are hard to maintain, understand, and visualise. GQM+Strategies™ [24] was developed to extend the Goal Question Metrics methodology by providing explicit support for the relation of software metrics measurement effort (e.g., measuring the impact that pair programming has on quality) to high-level business goals. However, the approach falls short in areas similar to the other methodologies reviewed; contribution links between goals are not quantified (i.e., assumed benefit), there is a fixed number of goal abstraction levels per diagram and there are no additional concepts that are typically included in GORE methodologies to place the goals into context (e.g., actors, conflicts, AND/OR refinement).

IV. METHODOLOGY

We propose that GRL goal graphs can be used to demonstrate strategic alignment by linking requirements as tasks (where the task is to implement the requirement) and business objectives as hard goals (where the hard goal brings about some business benefit) with contribution links (where the requirement is the means to the objective's end). The requirements should be abstracted (asking "why?") until they link to business objectives. We have used GRL's notation because it is part of the Z.151 international standard [3] and because its notation is well known (it originates from i*).

Soft goal elements (e.g., goals and visions from the Business Motivation Model) should not be defined in the goal graph for the purpose of demonstrating strategic alignment since their satisfaction is often immeasurable (if their satisfaction is possible at all); therefore, it is nonsensical to consider that a requirement may either partially or completely satisfy a goal or a vision. However, since objectives exist to quantify goals, and since goals exist in order to amplify the vision of the business [12], non-weighted traceability between an objective and its goals (and their related vision) should be maintained for posterity.

For our reference implementation, we have used the Volere requirements template to define the attributes of a requirement, primarily because it specifies a fit criterion field used for testing the requirement's satisfaction. An "estimated effort" field (specified in person-hours) could be added to the template so that cost-benefit analysis can be performed. Software implementation effort estimation methods such as COCOMO [25] could be useful in refining estimated values.

We define objectives using our modified GQM+Strategies formalisation template [26], as shown in Figure 1. Our modifications to the textual template attempt to improve integration with visual GRL diagrams through:

1. The addition of the scale concept from Planguage, which specifies the metric used for measurement. An objective's contribution to another is then given in terms of the second (parent) objective's scale.
2. The specification of the objective's activity attribute in the past tense, since objectives represent a desired outcome rather than an activity.
3. The removal of the constraints and relations fields since these can be expressed diagrammatically.
4. The addition of the author field so that newly proposed objectives can be identified and traced.

| Activity | Reduced |
|---|---|
| Object | TS&D Fabricated Structure Manufacturing |
| Focus | Lead Time |
| Magnitude | 3 months |
| Scale | Time in months required to have parts manufactured from the inception of a new engine |
| Timeframe | 1 year after system deployment |
| Scope | Transmissions Structures & Drives (TS&D) SCU |
| Author | John Smith (Component Engineer, TS&D) |

Figure 1: Example GQM+Strategies Formalisation

An objective is satisfied when the specified magnitude is achieved within the specified timeframe. The contribution links going toward an objective specify how that magnitude will be achieved (or exceeded). If the contributions of the child objectives additively amount to meet or exceed the objective's specified magnitude, then the satisfaction of the objective can be considered more likely than if not.

Figure 2 shows the GRL notation that is used to represent the requirements and objectives. Other notations could be used on the condition that they support the same concepts.

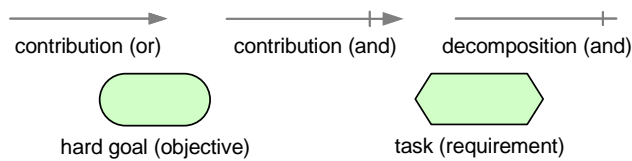

Figure 2: GRL Diagram Notation

In order to visualise the objectives specified with the GQM+Strategies template in a goal graph, we use GRL hard goal elements with the naming syntax: "Activity[Object Focus](magnitude)". We represent software requirements as tasks (i.e., the task of implementing the requirement) using the naming syntax: "{F/NF}[Requirement](Fit Criterion)", where "F/NF" is either Functional or Non-Functional, "Requirement" is a short headline version of the requirement description, and "Fit Criterion" is the short-hand version of the metric used to test the requirement's satisfaction.

A contribution link between a requirement and an objective specifies that the satisfaction of the requirement (tested by its fit criterion) will achieve some satisfaction of the objective, where the extent of the satisfaction is defined by the contribution specified by the link. A link between two objectives is similar, except for that the satisfaction of an objective is measured by its magnitude rather than by a fit criterion. An "OR" contribution specifies that if there are multiple "OR" links, a decision has to be made about which should be satisfied. An "AND" contribution specifies that all "AND" links are required for the objective to be satisfied. A decomposition link decomposes a requirement into a more specific requirement, much like SysML's "deriveReqt" link stereotype [8]. Figure 3 shows an example diagram produced by the methodology, which demonstrates the usage of the elements in Figure 2 to explore and visualise the strategic alignment of three high-level software requirements.

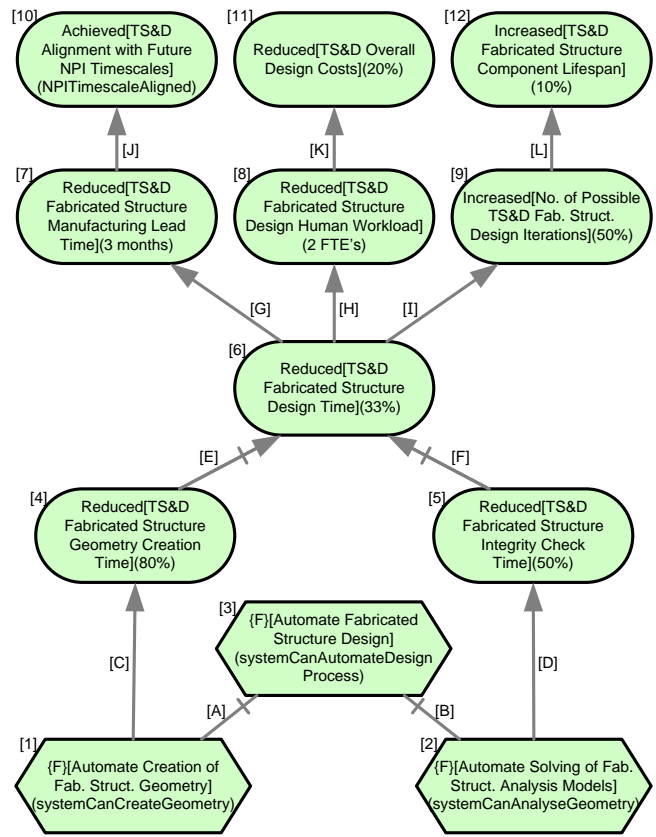

Figure 3: Example Strategic Alignment Diagram

The high-level software requirement (3) in Figure 3 is decomposed to two lower level software requirements (1 & 2) to represent the hierarchy of requirement abstraction. Such refinements through decomposition links will continue until the lowest level of requirements are represented. The decomposed requirements (1 & 2) then link to objectives (4 & 5) with contribution links in order to represent what those requirements hope to achieve. The contribution links (E & F) are of the "AND" type, since both objectives (4 & 5) are required if objective (6) is to be satisfied.

Table 1 shows a sample of the quantifications that complement the diagram. They have been separated out of the goal graph due to space constraints, but ordinarily would be annotated on the edges (connecting links) of the graph.

TABLE 1: QUANTIFIED CONTRIBUTIONS

| Link | [Contribution] [Activity] [Scale] | Confidence |
|---|---|---|
| C (1→4) | [80%] [Reduction] in [Geometry Creation Time] | 1 |
| D (2→5) | [50%] [Reduction] in [Integrity Check Time] | 0.75 |
| E (4→6) | [20%] [Reduction] in [Time Required to Design] | 1 |
| F (5→6) | [13%] [Reduction] in [Time Required to Design] | 0.75 |
| G (6→7) | [3 months] [Reduction] in [Manufacturing Lead Time] | 0.75 |

The quantified contributions in Table 1 tell us that objective (4) will be satisfied if requirement (1) is satisfied, since objective (4)'s required magnitude of satisfaction (80%) will be contributed by link (C) (80%). It is important to note that where percentages are used as contribution weights on links, this does not infer that a certain percentage of the objective's magnitude will be achieved (in this case, 80% of 80%). Instead, the focus of the objective (e.g., geometry creation time) will be affected by that percentage in the context of the activity (e.g., a reduction by 80%). Objective (4) is then abstracted until the benefits are expressed in terms of high-level business objectives, which disambiguates estimated business value by placing the quantifications into context (i.e., a large saving from a small cost may be less than a small saving from a large cost).

Confidence levels allow users to represent how sure they are that achieving the first objective (or requirement) affects the second objective by at least the specified contribution.

TABLE 2: CONFIDENCE LEVEL ENUMERATIONS

| Confidence | Description |
|---|---|
| 0.25 | Poor credibility, no supporting evidence or calculations, high doubt about capability |
| 0.5 | Average credibility, no evidence but reliable calculations, some doubt about capability |
| 0.75 | Great credibility, reliable secondary sources of evidence, small doubt about capability |
| 1 | Perfect credibility, multiple primary sources of evidence, no doubt about capability |

The confidence level concept is similar to that used by Gilb for impact estimation [23], so we base our confidence levels on a similar scale in Table 2. Basic confidence adjustment can be performed by multiplying contributions by their associated confidence level so that users are reminded of the impact confidence has on estimations. For example, when confidence levels are taken into consideration in Table 1, the satisfaction of requirement (1) still leads to the full satisfaction of objective (4). However, when confidence levels are considered for links (E & F), the satisfaction of objective (6) is in doubt, since (20*1) + (13*0.75) is less than the 33% required by the objective's magnitude attribute. Additional confidence levels can be applied to the user's estimations to represent how qualified that user is at providing estimations. For example, someone who has implemented similar systems should be able to provide more accurate estimations than someone who has not. The accuracy of previous estimates made by that person could also be considered in order to improve the reliability of the estimations (i.e., calibration of the confidence levels).

By traversing the quantified GRL goal graphs, the business value of a requirement can be calculated by the contribution it makes to business objectives. This calculation can be automated by using a graph traversal algorithm (e.g., depth-first search) to calculate how much a given requirement contributes to a business objective. This calculation could then be used to improve the outcome of requirements prioritisation methods such as the Analytics Hierarchy Process [27], since such pairwise methods depend on the practitioners understanding of a requirement's value.

It is important to note that software engineers and business analysts may not know the objectives (or the goals and visions, for that matter) at different levels of the business (i.e., the project, the business unit, the department, the overall business, etc.). Therefore, managers should work with stakeholders to define the business objectives before the requirements can be abstracted toward them. Indeed, it is likely that some software requirements will be abstracted toward business objectives that were not previously elicited.

Where typical goal abstraction (asking "why?") would allow a non-specific goal such as "improve the engine", this method requires the user to be specific in how the engine is to be improved by asking for the metric that will be affected, e.g., "component lifespan" from objective (12) in Figure 3. Users may resist quantifying benefits of requirements, especially for non-functional requirements where the subject may be intangible, however, Gilb has found that it has always been possible to do so in his experience (e.g., by polling customers to quantify customer satisfaction) and has provided guidance on doing so in [23]. Even if the magnitude cannot be elicited at first, providing a scale by which the objective's success will be measured improves the definition of the objective by reducing ambiguity.

We suggest that this methodology should be performed after the high-level requirements have been elicited, so that resources are not wasted eliciting lower level requirements that do not align well to business strategy.

Tool support (GoalViz) has been developed (free to download at [28]) to support the methodology through:

- Input support for the requirement and objective templates with prompt question generation.
- Automatic diagram drawing to focus the user on the methodology and data rather than the graph layout.
- Automatic evaluation and summarisation of chains of links to enable efficient understanding.
- Project libraries to facilitate learning about the estimated contributions made in previous projects to improve future quantification confidence levels.

- What-if analysis allowing comparison of outcomes for different inputs where there is some uncertainty.

## V. CONCLUSION AND FUTURE WORK

This paper's unique contribution is twofold. First, we have shown how quantified goal graphs can be used to visualise the alignment of software requirements to business objectives. We have shown that in order to demonstrate strategic alignment, a chain of objectives may contain different measurement scales, and, since strategic alignment is based on estimated benefit, confidence in the estimations should be made explicit. Secondly, we have shown how goal link contribution scores can be made testable by specifying them in terms of the estimated effect they will have on the parent goal's measurement scale. Our methodology not only facilitates disambiguation of a requirement's business value, but more importantly, it requires that the needs of the business (i.e., business objectives) are related to requirements to ensure that the software can add value to the business. Since the requirements are abstracted to several levels of objectives, the problem to be solved will have been defined even if the requirement was originally solution oriented.

Future work will evaluate this approach against the related work detailed in Section III within different industrial settings to examine its benefit in a range of domains. We also intend to evaluate integration with SysML [8] to improve traceability to the design that will realise the requirements.